\documentclass[aps,prl,superscriptaddress,showpacs,floatfix,nobibnotes]{revtex4}
\usepackage{epsfig}
\usepackage{epstopdf}

\usepackage{graphicx}
\usepackage{longtable}
\usepackage{CJK}
\usepackage{color}

\usepackage{mathptmx, courier, pifont}
\usepackage[scaled=0.92]{helvet}
\usepackage[T1]{fontenc}
\usepackage{textcomp}

\begin{document}

\begin{CJK*}{GB}{}

\title{Hidden Euclidean dynamical symmetry in the U($n+1$) vibron model}

\author{Yu Zhang }\email{dlzhangyu_physics@163.com}
\affiliation{Department of Physics, Liaoning Normal University,
Dalian 116029, P. R. China}

\author{Zi-Tong Wang}
\affiliation{Department of Physics, Liaoning Normal University,
Dalian 116029, P. R. China}

\author{Hong-Di Jiang }
\affiliation{Department of Physics, Liaoning Normal University,
Dalian 116029, P. R. China}

\author{Xin Chen }
\affiliation{Department of Physics, Liaoning Normal University,
Dalian 116029, P. R. China}

\date{\today}

\begin{abstract}
Based on the boson realization of the Euclidean algebras, it is found that the E($n$) dynamical symmetry (DS) may emerge at the critical point of the U($n$)-SO($n+1$) quantum phase transition. To justify this finding, we provide a detailed analysis of the critical dynamics in the U($n+1$) vibron model in both quantal and classical ways. It is further shown that the low-lying structure of $^{82}$Kr may serve as an excellent empirical realization of the E(5) DS in experiments.

\end{abstract}
\pacs{21.60.Fw, 03.65Fd, 05.30Jp}

\maketitle

\end{CJK*}

%\newpage
\begin{center}
\vskip.2cm\textbf{I. Introduction}
\end{center}\vskip.2cm

Dynamical symmetries (DSs) play an essential role in deeply understanding
the dynamical structures of quantum many-body systems. In general, A DS is supposed to occur
when the Hamiltonian of a system can be expressed as a combination of
the Casimir operators of a chain of Lie group $G\supset
G^\prime\supset G^{\prime\prime}\cdots$~\cite{IachelloBook2006}. DS of this type can be recognized by analyzing the associated algebraic
structure. The typical examples are
those associated with the interacting boson model (IBM)~\cite{IachelloBook87} and the U(4) vibron
model~\cite{IachelloBook95}. In the IBM, there are three typical DSs, U(5), SO(6) and SU(3), with their group generators
being all reduced from the U(6) ones composed of 36 bilinear products of the $s~d$ boson operators~\cite{IachelloBook87}.
Apart from exact DSs, approximate DSs are also suggested to exist in many-body systems too and yield some import symmetry-based concepts.
For instance, the partial dynamical symmetries~\cite{Leviatan1996,Leviatan2007,Ramos2009} and quasidynamical
symmetries~\cite{Rowe2004,Rowe2004II,Rowe2004III} have been found to occur in the IBM and other algebraic models.
The approximate DSs are usually hidden behind a complicate parameter relation of the Hamiltonian of a given system. One example is just the SU(3) approximate symmetry in the IBM~\cite{Bonatsos2010,Bonatsos2011}, which was found to be preserved along the trajectory in
the IBM parameter space close to the Alhassid-Whelan arc of regularity~\cite{Alhassid1991}.

There is another type of DS, called critical point symmetry (CPS)~\cite{Iachello2000}. As the first example, the E(5) CPS was proposed to describe spectra of nuclei around the critical point of the U(5)-SO(6) quantum phase transition (QPT)~\cite{Iachello2000}. This mode was built in the Bohr-Mottelson model~\cite{Bohrbook} by taking an infinite square well potential to simulate the mean-field structure at the critical point of the U(5)-SO(6) QPT in the IBM~\cite{IachelloBook87}. Then, the $d$-boson realization of the five-dimensional Euclidean DS was proposed~\cite{ZLPSD2014,Zhang2014} to give an algebraic description of the CPS mode, by which a much closer relation between the E(5) CPS and the U(5)-SO(6) QPT has been revealed. In addition, the E(5) CPS has been extended to the cases of $n=3$~\cite{Zhang2008} and $n=2$~\cite{Clark2006,Zhang2010} to describe other QPT systems. In view of the success of the Euclidean mode as a benchmark for critical structures in different systems, it is necessary to provide a general analysis of how an Euclidean DS shows up in quantum many-body systems.

In this work, we will propose a boson realization of the $n$-dimensional Euclidean algebra with $n=2l+1$ and $l=0,~1,~2,~\cdots$ and try to reveal the connection between the E($n$) DS and the U($n+1$) vibron model~\cite{IachelloBook2006}, which is used to describe the $2^l$-pole deformation of a many-body system. The famous examples of the U($n+1$) vibron model
are just the IBM (U(6)) describing quadrupole-deformation of nuclei~\cite{IachelloBook87} and the U(4) vibron model describing spectra of diatomic molecules corresponding to a dipole-deformed system~\cite{IachelloBook95}. They are also two important examples that will be discussed in this work.
\begin{center}
\vskip.2cm\textbf{II. The boson realization of the E($n$) algebra}
\end{center}\vskip.2cm

The $n$-dimensional Euclidean (E($n$)) space with $n=2l+1$ is generated by the coordinates $q_u^{(l)}$ $(u=0,~\pm1,~\pm2,~\pm l)$, to which the conjugate momenta are defined as $p_u^{(l)}=-i\frac{\partial}{\partial q_u^{(l)}}$. The associated $\mathrm{E}(n)$ group symmetry is then described as the invariance under
the translations and rotations in the E($n$) space~\cite{IachelloBook2006}. For convenience, the symbols $p_u(q_u)$ will be used instead of $p_u^{(l)}(q_u^{(l)})$ in the following discussion. The other conventions are
\begin{eqnarray}
&&\tilde{A}_u^{(\lambda)}=(-1)^{\lambda-u}A_{-u}^{(\lambda)},\\
&&(A^{(\lambda)})^2=\sum_uA_u^{(\lambda)}\tilde{A}_u^{(\lambda)}=(\tilde{A}^{(\lambda)})^2,\\
&&\tilde{A}^{(\lambda)}\cdot\tilde{A}^{(\lambda)}=\sum_u(-1)^u\tilde{A}_u^{(\lambda)}\tilde{A}_{-u}^{(\lambda)},\\
&&(\tilde{A}^{(\lambda)}\times\tilde{B}^{(\lambda^\prime)})_{u^{\prime\prime}}^{(\lambda^{\prime\prime})}=\sum_{uu^\prime}\langle\lambda u\lambda^\prime u^\prime\mid\lambda^{\prime\prime}u^{\prime\prime}\rangle\tilde{A}_u^{(\lambda)}\tilde{B}_{u^\prime}^{(\lambda^\prime)}\, ,
\end{eqnarray}
in which $\tilde{A}_u^{(\lambda)}(\tilde{B}^{(\lambda)})$ represents a spherical tensor of spin $\lambda$.
As we know, the $\mathrm{E}(n)$ group can be expressed as the semidirect product of $\mathrm{R}(n)$ and $\mathrm{SO}(n)$, namely \begin{eqnarray}\mathrm{E}(n)=\mathrm{R}(n)\otimes_\mathrm{s}\mathrm{SO}(n)\, ,
\end{eqnarray}
where $\mathrm{R}(n)$ represents the $n$-dimensional translation group generated by $i\tilde{p}_u$ and $\mathrm{SO}(n)$ denotes the $n$-dimensional rotation group generated by $\hat{T}_u^{(k)}\equiv i(q\times \tilde{p})_u^{(k)}$ with $k=1,~3,\cdots,~2l-1$.
The kinetic energy term $p^2=\sum_u\tilde{p}_up_u$ is shown to be an invariant quantity of the $\mathrm{E}(n)$ group with
$[i\tilde{p}_u,~p^2]=0$ and $[\hat{T}_u^{(k)},~p^2]=0$.
Considering the conservation of angular momentum, the $\mathrm{E}(n)$ DS in a many-body system may be characterized by the group chain
\begin{equation}\label{GE(2l+1)}
\mathrm{E}(n)\supset \mathrm{SO}(n)\supset \mathrm{SO(3)}\,
\end{equation}
with the angular momentum group SO($3$) being generated by $\hat{T}_u^{(1)}$.

To build a boson realization of the E($n$) algebra, one can define the boson operators of spin $l$ by
\begin{eqnarray}\label{bb}
b_u^{l\dag}=\frac{1}{\sqrt{2}}[q_u-i\tilde{p}_u],~~b_u^{l}=\frac{1}{\sqrt{2}}[\tilde{q}_u+ip_u]\, ,
\end{eqnarray}
where $l=0,~1,~2,\cdots$ stand for the bosons, $s,~p,~d,\cdots$. The $l=2$ case in the definition (\ref{bb}) was given in \cite{Castanos1979}.
With this definition, it is easy to prove
the boson commutation relation, $[b_{u^\prime}^{l^\prime},~b_u^{l\dag}]=\delta_{u^\prime u}\delta_{l^\prime l}$.
The $n(n+1)/2$ generators of the $\mathrm{E}(n)$ Lie algebra are then rewritten as
\begin{eqnarray}\label{gE1}
\hat{\Lambda}_u^{(l)}&\equiv&i\tilde{p}_u=\frac{1}{\sqrt{2}}[\tilde{b}_u^{l}-b_u^{l\dag}],\\ \label{gE2}
\hat{T}_u^{(k)}&\equiv&i(q\times\tilde{p})_u^{(k)}=(b^{l\dag}\times\tilde{b}^l)_u^{(k)},~~~k=1,~3,~\cdots,~2l-1\, .
\end{eqnarray}
The 2nd-order Casimir operator of the
$\mathrm{E}(n)$ group is given by
\begin{eqnarray}\label{C2}
\hat{C}_2[\mathrm{E}(n)]\equiv p^2=\hat{n}_{b^l} + \frac{n}{2}-
\frac{1}{2}(-)^l\left(\hat{P}^{\dag}_{l} + \hat{P}_{l} \right)\,
\end{eqnarray}
with
\begin{eqnarray}\label{nP}
\hat{n}_{b^l}=\sum_{u} b_u^{l\dag} b_{u}^{l},~~~~\hat{P}_{l} =
\sum_{u} (-)^{u} \tilde{b}_{u}^l \tilde{b}_{-u}^l,~~~~\hat{P}^{\dag}_{l}=(\hat{P}_l)^{\dag}\, .
\end{eqnarray}
It can be proved that these generators satisfy the
commutation relations:
\begin{eqnarray}\label{LL}
&&[\hat{\Lambda}_u^{(l)},\hat{\Lambda}_v^{l}]=0~,\\ \label{TL}
&&[\hat{T}_u^{(k)},\hat{\Lambda}_v^{(l)}]=-\sqrt{\frac{(2k+1)}{2l+1}}\langle k
u lv|lu+v\rangle \hat{\Lambda}_{u+v}^{(l)}~,\\ \nonumber \label{TT}
&&[\hat{T}_u^{(k)},\hat{T}_{\bar{u}}^{(\bar{k})}]=-2\sqrt{(2k+1)(2\bar{k}+1)}\sum_{\lambda=\mathrm{odd}}
\left\{\begin{array}{cc} k,\bar{k},\lambda \\
l,~l,~l\,\end{array}\right\}\, \\
&&~~~~~~~~~~~~~~~~~~~~~~~~~\times\langle k u \bar{k}
\bar{u}|\lambda u+\bar{u}\rangle~\hat{T}_{u+\bar{u}}^{(\lambda)}\, ~,
\end{eqnarray}
and
\begin{equation}
\Big[\hat{\Lambda}_u^{(l)},~\hat{C}_2[\mathrm{E}(n)]\Big]=\Big[\hat{T}_u^{(k)},~\hat{C}_2[\mathrm{E}(n)]\Big]=0\,
.
\end{equation}
Clearly, $n$ is an odd number with $n=2l+1$ and $\mathrm{E}(n)$ is a non-compact Lie group.

\begin{center}
\vskip.2cm\textbf{III. The U($n$+1) algebra and group contraction}
\end{center}\vskip.2cm

The U($n+1$) vibron model with $n=2l+1$ can be applied to describe the $2^l$-pole deformation dynamics of a many-body system~\cite{IachelloBook2006}. Hamiltonian in the U($n+1$) vibron model is constructed from two kinds of boson operators: the scalar $s$-boson and the $l$-rank tensor $b^l$-boson. The $(n+1)^2$ bilinear operators
\begin{eqnarray}
s^\dag s,~~~~s^\dag b_u^{l},~~~~b_u^{l\dag}s,~~~~b_u^{l\dag}b_v^l~~~~u,~v=-l,~-l+1,~\cdots,~l\,
\end{eqnarray}
generate the maximal dynamical symmetry group, U($n+1$). It can be proved that one of its subgroups is just U($n$) generated by
\begin{eqnarray}
\hat{B}_q^{(k)}=(b^{l\dag}\times\tilde{b}^l)_q^{(k)},~~~~k=0,~1,~2,~\cdots,~2l\,
\end{eqnarray} with the algebraic relation
\begin{eqnarray}
&&[\hat{B}_u^{(k)},\hat{B}_{\bar{u}}^{(\bar{k})}]=\sqrt{(2k+1)(2\bar{k}+1)}\sum_{\lambda}
\left\{\begin{array}{cc} k,\bar{k},\lambda \\ \label{RR}
l,~l,~l\,\end{array}\right\}\, \\ \nonumber
&&~~~~~~~~~~~~~~~~~~\times\Big((-)^{\lambda}-(-)^{k+\bar{k}}\Big)\langle k u \bar{k}
\bar{u}|\lambda u+\bar{u}\rangle\hat{B}_{u+\bar{u}}^{(\lambda)}\,
\end{eqnarray}
and another subgroup is SO($n+1$) generated by
\begin{eqnarray}\label{QT}
\hat{Q}_u=(s^\dag\times\tilde{b}^l+b^{l\dag}\times \tilde{s})_u^{(l)},~~~~\hat{T}_q^{(k)}=\hat{B}_q^{(k)},~~~~k=\mathrm{odd}\,
\end{eqnarray}
with the algebraic relation
\begin{eqnarray} \label{QQ}
&&[\hat{Q}_u,~\hat{Q}_v]=2\sum_{k}\langle lulv|ku+v\rangle\hat{T}_{u+v}^{(k)}~,\\ \label{TQ}
&&[\hat{T}_u^{(k)},\hat{Q}_v]=-\sqrt{\frac{2k+1}{2l+1}}\langle kulv|lu+v\rangle\hat{Q}_{u+v}\, .
\end{eqnarray}
Note that the commutation relation $[\hat{T}_u^{(k)},\hat{T}_{\bar{u}}^{(\bar{k})}]$ is the same as that given in (\ref{TT}).
Clearly, SO($n$) is the common subgroup of U($n$) and SO($n+1$).
Two typical DSs in
the U($n+1$) vibron model are accordingly characterized by the group chains~\cite{IachelloBook2006}
\begin{eqnarray}\label{ulimit}
&&\mathrm{U}(n+1) \supset \mathrm{U}(n) \supset \mathrm{SO}(n) \supset \mathrm{SO(3)}\, ,\\ \label{olimit}
&&\mathrm{U}(n+1) \supset \mathrm{SO}(n+1) \supset \mathrm{SO}(n) \supset
\mathrm{SO(3)}\, \, .
\end{eqnarray}
The relevant Casimir operators are defined by
\begin{eqnarray}\label{c1}
&&\hat{C}_1[\mathrm{U}(n)]=\hat{n}_{b^l},\\ \label{c2}
&&\hat{C}_2[\mathrm{U}(n)]=\hat{n}_{b^l}(\hat{n}_{b^l}+n-1),\\ \label{c3}
&&\hat{C}_2[\mathrm{SO}(n+1)]=(-1)^l\hat{Q}\cdot\hat{Q}+2\sum_{k}\hat{T}^{(k)}\cdot\hat{T}^{(k)},\\ \label{c4}
&&\hat{C}_2[\mathrm{SO}(n)]=2\sum_{k}\hat{T}^{(k)}\cdot\hat{T}^{(k)},\\ \label{c5}
&&\hat{C}_2[\mathrm{SO}(3)]=\frac{l(l+1)(2l+1)}{3}\hat{T}^{(1)}\cdot\hat{T}^{(1)}\, ,
\end{eqnarray}
where the operators $\hat{n}_{b^l}$, $\hat{Q}_u$ and $\hat{T}_u^{(k)}$ are those defined in (\ref{nP}) and (\ref{QT}).
The corresponding eigenvalues can be expressed as
\begin{eqnarray}\label{e1}
&&\langle\hat{C}_1[\mathrm{U}(n)]\rangle={n}_{b^l},\\ \label{e2}
&&\langle\hat{C}_2[\mathrm{U}(n)]\rangle=n_{b^l}(n_{b^l}+n-1),\\ \label{e3}
&&\langle\hat{C}_2[\mathrm{SO}(n+1)]\rangle=\sigma(\sigma+n-1),\\ \label{e4}
&&\langle\hat{C}_2[\mathrm{SO}(n)]\rangle=\omega(\omega+n-2),\\ \label{e5}
&&\langle\hat{C}_2[\mathrm{SO}(3)]\rangle=L(L+1)\,
\end{eqnarray}
with the quantum numbers ${n}_{b^l},~\sigma,~\omega,$ and $L$ being used to signify
the irreducible representations of U($n$), SO($n+1$), SO($n$) and SO(3), respectively~\cite{IachelloBook2006}.

One can find hat the SO($n+1$) algebraic structure is very similar to the E($n$) one.
Both of them have the SO($n$) algebra as their subalgebra.
If the generators are rescaled by~\cite{Bonatsos2011}
\begin{eqnarray}\hat{q}_u=\frac{1}{\sqrt{C_2[\sigma]}}\hat{Q}_u\,
\end{eqnarray}
with $C_2[\sigma]=\sigma(\sigma+2l)$,
the algebraic relation shown in (\ref{TQ}) and (\ref{QQ}) will be changed into
\begin{eqnarray}\label{tq}
&&[\hat{T}_u^{(k)},~\hat{q}_v]=-\sqrt{\frac{2k+1}{2l+1}}\langle kulv|lu+v\rangle\hat{q}_{u+v},\\ \label{qq2}
&&[\hat{q}_u,~\hat{q}_v]=2\sum_{k}\langle lulv|ku+v\rangle\frac{1}{C_2[\sigma]}\hat{T}_{u+v}^{(k)}\, .
\end{eqnarray}
In the $\sigma\rightarrow\infty$ limit, it is given by $[\hat{q}_u,~\hat{q}_v]\simeq0$ for small
$\omega$ cases, in which the expectation values $\langle \hat{T}{(k)}\rangle$ should be small. By this
procedure, one may get the same commutation relations as those shown in (\ref{LL})-(\ref{TT}) with the correspondence $\hat{\Lambda}\rightarrow \hat{q}$. It means that the $\mathrm{SO}(n+1)\rightarrow \mathrm{E}(n)$ contraction may happen in the $\sigma\rightarrow\infty$ limit. Such a group contraction for $n=5$
has been previously discussed in \cite{Bonatsos2011} and earlier in \cite{Meyer1979}. The present result is a direct generalization of the $n=5$ case. Another example of group contraction worth mentioning is $\mathrm{SU}(3)\rightarrow \mathrm{R}(5)\otimes_s\mathrm{SO}(3)$, which can be achieved via the similar procedure~\cite{Bonatsos2011,Kota2020}. This group contraction provides the theoretical basis to construct the SU(3) image of the triaxial rotor dynamics~\cite{Leschber1987,Castanos1988}, which offers a microscopic way of understanding collective rotations in triaxial nuclei based on the SU(3) shell model~\cite{Draayer1983,Draayer1989}. Different aspects of the SU(3) symmetry in nuclei can be read from \cite{Kota2020}.

\begin{center}
\vskip.2cm\textbf{IV. The emerged E($n$) DS in the U($n$)-SO($n+1$) QPT}
\end{center}\vskip.2cm

In contrast to the group contraction, we hope to emphasize in this work another way in which the E($n$) symmetry can dynamically emerge from the U($n+1$) vibron model. As discussed above, U($n$) and SO($n+1$) are two typical dynamical symmetry limits in the vibron model. Hamiltonian for either of them can be written as a linear combination of the Casimir operators of the corresponding group chain so that the eigenvalues and eigenvectors can be expressed in an analytical way (see Eq.~(\ref{e1})-(\ref{e5})). In general, no analytical solutions can be achieved in the cases of symmetry-mixing. Nevertheless, beautiful algebraic solutions of the Hamiltonian mixing the U($n$) and O($n+1$) DSs have been obtained in \cite{Pan1998,Pan2002} using the Bethe ansatz within an infinite-dimensional Lie algebra.

\begin{center}
\vskip.2cm\textbf{A. The transitional Hamiltonian}
\end{center}\vskip.2cm
To discuss a general situation in the U($n+1$) vibron model, we adopt here a schematic Hamiltonian
\begin{eqnarray}\label{H}
\hat{H}&=&\varepsilon\Big[(1-\eta)\hat{n}_{b^l}-\frac{\eta}{4N}(-1)^l\hat{Q}\cdot\hat{Q}\Big]\\ \nonumber
&=&\varepsilon\Big[(1-\eta)\hat{C}_1[\mathrm{U}(n)]-\frac{\eta}{4N}\hat{C}_2[\mathrm{SO}(n+1)]\\ \nonumber
&~&+\frac{\eta}{4N}\hat{C}_2[\mathrm{SO}(n)]\Big]\, ,
\end{eqnarray}
where $N=n_s+n_{b^l}$ represents the total boson number of the system and $\varepsilon$ is a scale parameter to be set
with $\varepsilon=1$ in the discussions.
It is apparent that the system is in the U($n$) DS when the control parameter $\eta=0$ and changed into the
SO($n+1$) DS when $\eta=1$. By varying $\eta\in$[0, 1], the Hamiltonian (\ref{H}) describes a
transitional situation in between the U($n$) and SO($n+1$) symmetry limits. In addition to U($n$) and SO($n+1$), there may exist other DSs in the U($n+1$)
vibron model for $n\geq5$. For example, the SU(3) DS will be involved in the $n=5$ case~\cite{IachelloBook87}. Nonetheless, it is sufficient to discuss the cases involving the U($n$) and SO($n+1$) DSs for the present purpose. The reason is that the $\mathrm{SO}(n)\supset \mathrm{SO}(3)$ DS (see Eq.~(\ref{GE(2l+1)})) will be conserved in the vibron model only in the U($n$) and SO($n+1$) limits or their mixing. Therefore, one only need to analyze the transitional Hamiltonian like that given in (\ref{H}) to reveal the underlying E($n$) DS in the U($n+1$) vibron model.

\begin{center}
\vskip.2cm\textbf{B. Quantal analysis}
\end{center}\vskip.2cm

As is known~\cite{Bonatsos2011}, if a system has an underlying symmetry of the group G,
the corresponding Hamiltonian
should commute with all the generators of the group G.
To identify the underlying E($n$) DS in the  parameter space of the vibron model, we
examine the commutation relations between the generators of the E($n$) group defined in (\ref{gE1})-(\ref{gE2}) and the
Hamiltonian (\ref{H}).
First of all, one can derive that
\begin{eqnarray}\label{commutation}
&&[\hat{T}_u^{(k)},~\hat{n}_{b^l}]=0,\\
&&[\hat{\Lambda}_u^{(l)},~\hat{n}_{b^l}]=\frac{1}{\sqrt{2}}(\tilde{b}_u^l+b_u^{l\dag}),\\
&&[\hat{\Lambda}_u^{(l)},~\hat{Q}_v]=\frac{(-)^{l-u}}{\sqrt{2}}\delta_{u,-v}(s+s^\dag)\, .
\end{eqnarray}
With the commutators, it is easy to prove
$[\hat{T}_u^{(k)},~\hat{H}]=0$. This point actually reflects the fact that $\hat{T}_u^{(k)}$
as the generators of U($n$) and SO($n+1$) should commute with
their Casimir operators.
Furthermore, one can derive in the $\langle \hat{n}_{b^l}/N\rangle\rightarrow0$ limit that
\begin{eqnarray}\nonumber
[\hat{\Lambda}_u^{(l)},~\hat{H}]&=&\frac{\sqrt{2}}{2}(1-\eta)(\tilde{b}_u^l+b_u^{l\dag})\\ \nonumber
&-&\frac{\sqrt{2}\eta}{4N}
\Big(s^\dag s^\dag\tilde{b}_u^l+ssb_u^{l\dag}+s^\dag s\tilde{b}_u^l+ss^\dag b_u^{l\dag}\Big)\\ \label{CC}
&\approx&\mid_{\langle\hat{n}_{b^l}/N\rangle\rightarrow0}\frac{\sqrt{2}}{2}(1-2\eta)(\tilde{b}_u^l+b_u^{l\dag})\, .
\end{eqnarray}
In the derivation, we have used the replacements
\begin{eqnarray}\nonumber
s^\dag(s)&\rightarrow&\sqrt{n_s+1}(\sqrt{n_s})\\
&=&\sqrt{N-n_{b^l}+1}(\sqrt{N-n_{b^l}})\simeq\sqrt{N}\, ,
\end{eqnarray}
which should be well satisfied in the $\langle \hat{n}_{b^l}/N\rangle\rightarrow0$ limit due to
$N=n_{b^l}+n_s$.
It is clear that the commutator given in (\ref{CC}) will vanish at $\eta=1/2$. Therefore, we conclude that the vibron Hamiltonian (\ref{H}) at
the parameter point $\eta=1/2$ is invariant under the E($n$) group
transformations when $\langle \hat{n}_{b^l}/N\rangle\rightarrow0$. In other words,
the E($n$) DS will occur in the U($n$)-SO($n+1$) transitional region under this approximate condition.

To examine the required condition for the E($n$) DS, we take $n=3$ and $n=5$ to represent the examples of $l=\mathrm{odd}$ and $l=\mathrm{even}$, respectively. For $n=5$, the U($n+1$) vibron model is reduced to the IBM (U(6))~\cite{IachelloBook87}.
Accordingly, the Hamiltonian in (\ref{H}) can be used to describe nuclear structural evolution
from the spherical vibration (U(5) DS) to $\gamma$-unstable rotation (SO(6) DS). To solve this
Hamiltonian, one can diagonalize it within the U(5) basis of the IBM
\begin{eqnarray}\label{u5}
\mid\phi\rangle_{\mathrm{U}(5)}=\mid N~n_d~\tau~\Delta~L\rangle\, .
\end{eqnarray}
Here, $N,~n_d,~\tau$ and $L$ represent the quantum numbers for U(6), U(5), SO(5) and SO(3), while $\Delta$
denotes the additional quantum number in the reduction SO(5)$\supset$SO(3)~\cite{IachelloBook87}.
With the solved wavefunctions, one can calculate the expectation value $\rho(\eta)=\langle\phi \mid\hat{n}_d/N\mid\phi\rangle$ for any given state $\mid\phi\rangle$ to check the condition $\langle \hat{n}_{b^l}/N\rangle\rightarrow0$.
If taking $n=3$, the U(4) vibron model for molecular spectra is obtained with the Hamiltonian in (\ref{H}) being used to describe the U(3)-SO(4) transition~\cite{IachelloBook95}. Similarly, one can worked out the expectation value $\rho(\eta)$ through diagonalizing the transitional Hamiltonian in the U(3) basis
\begin{eqnarray}\label{u3}
\mid\phi\rangle_{\mathrm{U}(3)}=\mid N~n_p~L\rangle\, ,
\end{eqnarray}
where $N$, $n_p$ and $L$ represent the quantum numbers for U(4), U(3) and SO(3), respectively~\cite{IachelloBook95}. In (\ref{u5}) and (\ref{u3}), the angular momentum projection, $M$, has been ignored for convenience.

\begin{figure}
\begin{center}
\includegraphics[scale=0.4]{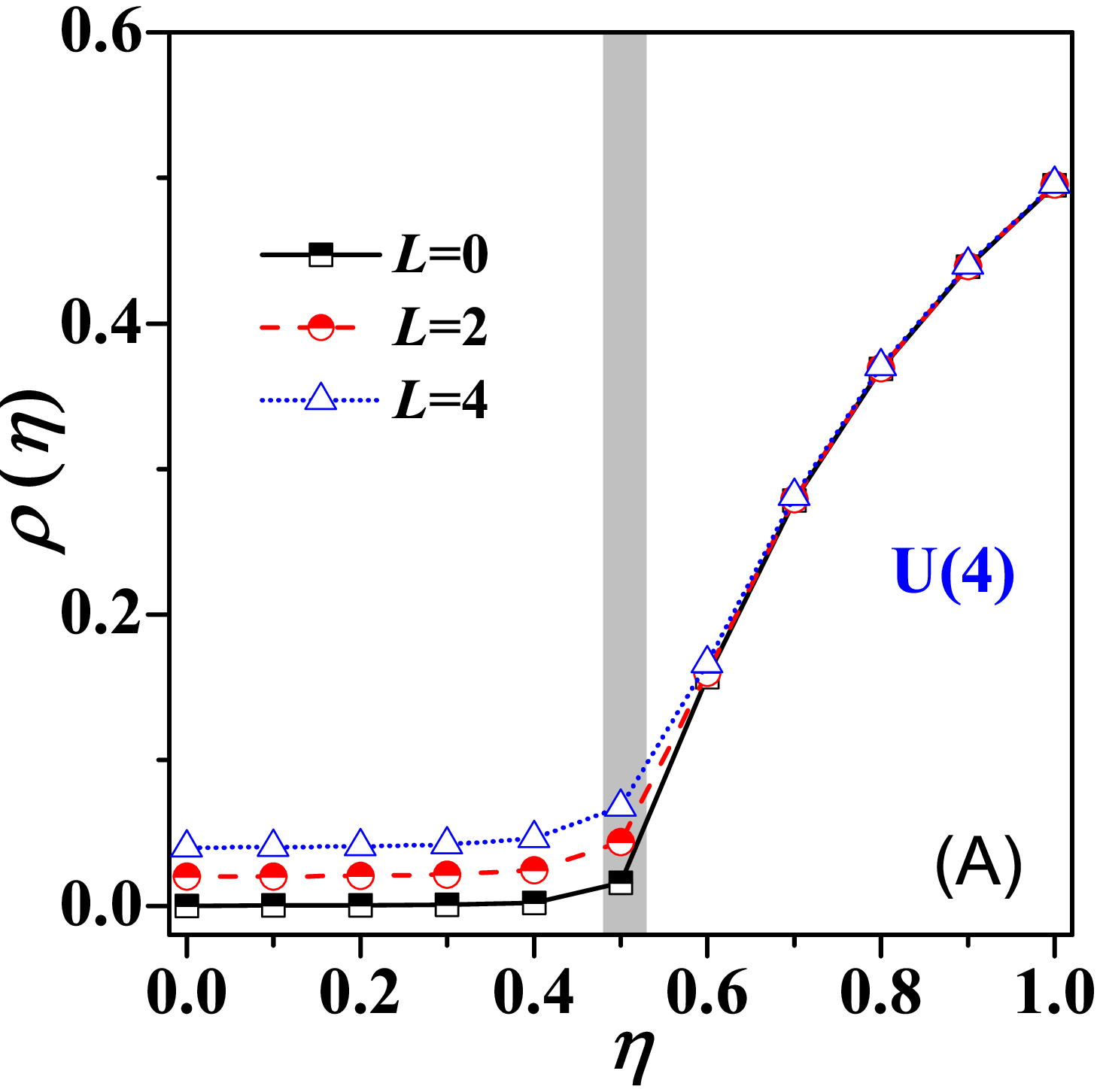}
\includegraphics[scale=0.4]{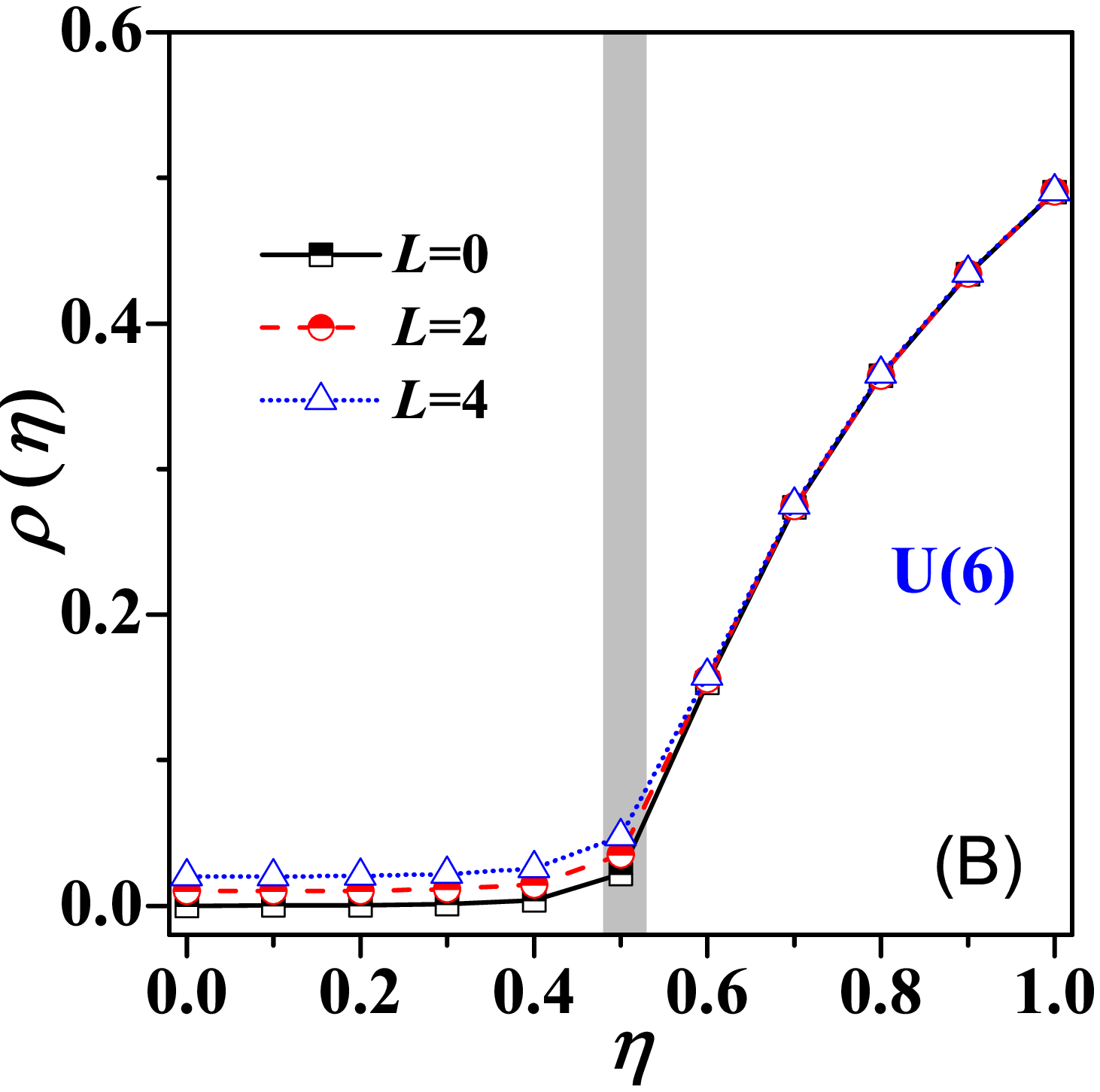}
\caption{(A)The order parameter $\rho(\eta)=\langle\hat{n}_b^l\rangle/N$ changes as a function of $\eta$ with the results solved from the U(4) model with $N=100$ for the lowest states of $L=0,~2,~4$. (B) The same as in (A) but for those solved from the U(6) model.  \label{F1}}
\end{center}
\end{figure}

In Fig.~\ref{F1}, the evolutions of $\rho(\eta)$ for the lowest states with $L=0,2,4$ are shown for both the U(4) and U(6) models. The total boson numbers in the two models are both taken as $N=100$. One can find from Fig.~\ref{F1} that $\rho(\eta)$ in the two models exhibit nearly the same evolutional behaviors. Specifically, the values of $\rho$ as a function of $\eta$ remain with $\rho\sim0$ for $\eta\in[0,~1/2]$ and rapidly increase when $\eta>1/2$. It is thus justified that the condition $\langle \hat{n}_{b^l}/N\rangle\rightarrow0$ are indeed well satisfied in these cases.
Moreover, the larger the boson number $N$, the better the approximation $\langle \hat{n}_{b^l}/N\rangle\rightarrow0$. If $N\rightarrow\infty$, the condition $\langle \hat{n}_{b^l}/N\rangle=0$ at $\eta=1/2$ will be exactly achieved for the ground state, which will be discussed later. It means that E($n$) could be an exact ground-state DS in the vibron model in the large-$N$ limit. In addition, the sudden enhancements in $\rho(\eta)$ as shown in Fig.~\ref{F1} actually manifests that this is a precursor of the U($n$)-SO($n+1$) QPT defined in the large-$N$ limit. According to the Ehrenfest classification, such an QPT is suggested to be second order. That is, the ground state energy $E_g$ and its first derivative, $\frac{\partial E_g}{\partial\eta}$, are both continuous at the transitional point but the second derivative, $\frac{\partial^2 E_g}{\partial\eta^2}$, is discontinuous~\cite{IachelloBook87}. Based on the Hellmann-Feynman theorem, $\frac{\partial E_n}{\partial\eta}=\langle\frac{\partial H}{\partial\eta}\rangle_n$, one can further derive that
\begin{eqnarray}\label{order}
\frac{\partial e_g}{\partial\eta}=\frac{1}{\eta}(e_g-\rho(\eta)_g),~~~~\frac{\partial^2 e_g}{\partial\eta^2}=-\frac{1}{\eta}\frac{\partial\rho(\eta)_g}{\partial\eta}\, ,
\end{eqnarray}
where $e_g\equiv E_g/N$ represents the ground state energy per boson.
It is suggested that the quantity $\rho(\eta)_g=\langle\hat{n}_b^l/N\rangle_g$ may serve as a quantum order parameter~\cite{Iachello2004}
for the U($n$)-SO($n+1$) QPT.
As observed from Fig.~\ref{F1}, one can locate the critical point of this QPT at $\eta_c=1/2$ according to the transitional behaviors of $\rho$. It means that the E($n$) DS may simultaneously play a role of critical point symmetry in the U($n+1$) vibron model~\cite{Iachello2000,Caprio2007}.

\begin{center}
\vskip.2cm\textbf{C. Classical analysis}
\end{center}\vskip.2cm

In the following, we will give a classical analysis of the transitional Hamiltonian (\ref{H}) to further identify the hidden E($n$) DS in 
the classical system.
In the definitions (\ref{bb}), one can extract the scalar boson operator in terms of
one-dimensional coordinate and momentum~\cite{Castanos1979}
\begin{eqnarray}\label{sboson}
s^{\dag}=\frac{1}{\sqrt{2}}[q_s-ip_s],~~~~s=\frac{1}{\sqrt{2}}[q_s+ip_s]\, .
\end{eqnarray}
The inverse transformations
\begin{eqnarray}\label{ss}
q_s=\frac{1}{\sqrt{2}}[s^\dag+s],~~~~p_s=\frac{i}{\sqrt{2}}[s^\dag-s]\,
\end{eqnarray}
indicate $q_s=\tilde{q}_s$ and $p_s=\tilde{p}_s$.
Then, the transitional Hamiltonian (\ref{H}) in the classical limit can be expressed as
\begin{eqnarray}\label{Hpq}
H(q,p,q_s,p_s)&=&(1-\eta)\frac{1}{2}\sum_u(q_u-i\tilde{p}_u)(\tilde{q}_u+ip_u)\\ \nonumber
&-&\frac{\eta}{4N}(-1)^l
Q(q,p,q_s,p_s)\cdot Q(q,p,q_s,p_s)\,
\end{eqnarray}
with
\begin{eqnarray}
Q=\frac{1}{2}\Big[(q_u-i\tilde{p}_u)(q_s+ip_s)+(q_s-ip_s)(q_u+i\tilde{p}_u))\Big]\, .
\end{eqnarray}
For classical system, the operators $q_u,~p_u,~q_s,~p_s$ become the ordinary coordinates and momenta.
Then, the classical Hamiltonian is reduced to
\begin{eqnarray}\label{CH}
H(q,p,q_s,p_s)&=&\frac{1-\eta}{2}(q^2+p^2)\\ \nonumber
&-&\frac{\eta}{4N}\Big[q^2q_s^2+p^2p_s^2+2p_sq_s\sum_u(p_uq_u)\Big]\, .
\end{eqnarray}
Accordingly, the boson number conservation condition $N=n_s+n_b$ leads to the classical constraint
\begin{eqnarray}\label{condition}
\frac{1}{2}(q^2+p^2+q_s^2+p_s^2)=N\, .
\end{eqnarray}
With the condition $n_b/N\rightarrow0$ indicating $s^\dag(s)\simeq\sqrt{N}$, one
may get $p_s=0$ and $q_s=\sqrt{2N}$ (see Eq.~(\ref{ss})). Then,
the classical Hamiltonian (\ref{CH}) is further reduced to \begin{eqnarray}\label{HF}
H(q,p,q_s=\sqrt{2N},p_s=0)\simeq\frac{1-2\eta}{2}q^2+\frac{1-\eta}{2}p^2\, .
\end{eqnarray}
It is easy to deduce from (\ref{HF}) that the Hamiltonian describes an $n$-dimensional harmonic oscillator with
\begin{eqnarray}
H\mid_{\eta=0}=\frac{1}{2}(q^2+p^2)\,
\end{eqnarray}
at $\eta=0$ and describes an E($n$) DS system with
\begin{eqnarray}
H\mid_{\eta=\frac{1}{2}}=\frac{1}{4}p^2=\frac{1}{4}\hat{C}[\mathrm{E}(n)]\,
\end{eqnarray}
at $\eta=1/2$.
Obviously, the E($n$) DS in its classical limit just corresponds to a free Hamiltonian~\cite{Caprio2007}.
It is thus justified that the E($n$) DS may also hide in the classical system described by the same Hamiltonian under the same condition.
Note that the approximation condition $n_b/N\rightarrow0$ is not satisfied for $\eta>1/2$, which means that one cannot derive the classical limit of the SO($n+1$) DS from (\ref{HF}) at $\eta=1$.

In the quantal analysis, the condition $n_b/N\rightarrow0$ has been checked in a
numerical way. In the following, we will identify this condition at mean-field level.
To do that, one needs to work out the scaled classical potential, which can be derived from the Hamiltonian (\ref{Hpq}) and given as
\begin{eqnarray}\label{CV}
V(q,q_s)&\equiv&\frac{1}{N}H(q,p,q_s,p_s)\mid_{p=0,p_s=0}\\ \nonumber
&=&\frac{1-\eta}{2N}q^2-\frac{\eta}{4N^2}q^2q_s^2\, .
\end{eqnarray}
The truly classical limit is obtained for $N\rightarrow\infty$ with the inverse of the boson number $1/N$ playing a role of $\hbar$~\cite{Alhassid1991,Alhassid1991II}.
So, we rescale here the coordinates with $\bar{q}=q/\sqrt{N}$ and $\bar{q_s}=q_s/\sqrt{N}$. The constraint condition in (\ref{condition}) is now changed into
\begin{eqnarray}
\frac{1}{2}(\bar{q}^2+\bar{q}_s^2)=1\, ,
\end{eqnarray}
by which the scaled potential is further reduced to
\begin{eqnarray}\label{CV2}
V(\bar{q})=(\frac{1}{2}-\eta)\bar{q}^2+\frac{\eta}{4}\bar{q}^4\,
\end{eqnarray}
with $\bar{q}^2\leq2$. To
see the deformation dependence of the potential, one should transform it into
the intrinsic coordinate system through a rotation with Euler angles $\Omega$,
\begin{eqnarray}\bar{q}_u=\sum_vD_{u,v}^{(l)}(\Omega)\bar{\beta}_{v}\, ,
\end{eqnarray} where $D^{(l)}$ are the Wigner matrices of spin $l$ and $\bar{\beta}_v$ represent the intrinsic coordinates.
But this transformation may generate $\bar{q}^2=\bar{\beta}^2$ and $\bar{q}^4=\bar{\beta}^4$, which leaves the potential function form unchanged with
\begin{eqnarray}\label{VB}
V(\bar{\beta})=(\frac{1}{2}-\eta)\bar{\beta}^2+\frac{\eta}{4}\bar{\beta}^4\, .
\end{eqnarray}
It means that the classical potential is only a function of the intrinsic "deformation" measured by $\bar{\beta}=\sqrt{\bar{\beta}^2}$.
By further minimizing the potential with respect to $\bar{\beta}$, one can get the ground state energy per boson
\begin{equation}
e_g(\eta)\equiv V(\eta,\bar{\beta}_e) =  \left\{ \begin{array}{cc} 0 \, , & ~~~~~ 0\leq\eta\leq 1/2 \, ,  \\
-\frac{(1-2\eta)^2}{4\eta} \, , &  \eta > 1/2 \,  ,\end{array} \right.
\end{equation}
where $\bar{\beta}_e$ denotes the optimal value of $\bar{\beta}$, namely the ground state deformation.
According to the Ehrenfest classification, one can prove that there exists a second-order QPT occurring at the parameter point $\eta=1/2$, i.e. the U($n$)-SO($n+1$) QPT. While, the E($n$) DS may occur at the same parameter point. It is thus confirmed that the Euclidean symmetry can indeed serve as a critical point symmetry of the U($n$)-SO($n+1$) QPT.

\begin{figure}
\begin{center}
\includegraphics[scale=0.45]{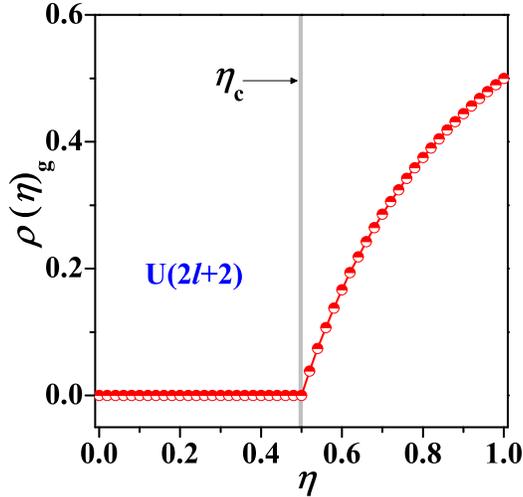}
\caption{The evolution of $\rho(\eta)_g=\langle n_b^l/N\rangle_g$ as a function of $\eta$ with the
analytical expression given in (\ref{rhog}). \label{F2}}
\end{center}
\end{figure}

For the U($n$)-SO($n+1$) QPT, the ground state deformation $\bar{\beta}_e$ can be taken as the classic order parameter~\cite{Iachello2004}. Its values can be derived as
\begin{equation}
\bar{\beta}_e(\eta) =  \left\{ \begin{array}{cc} 0 \, , & ~~~~~ 0\leq\eta\leq 1/2 \, ,  \\
\sqrt{2(\eta-1/2)/\eta} \, , &  \eta > 1/2 \,  .\end{array} \right.
\end{equation}
Then, one can extract the critical exponent $u=1/2$ by expanding the order parameter around the critical point $\eta_c$~\cite{Zhang2008}
with
\begin{eqnarray}
[\bar{\beta}_e(\eta)-\bar{\beta}_e(\eta_c)]\propto(\eta-\eta_c)^u\, .
\end{eqnarray}
Based on the relation given in (\ref{order}), one can further work out $\rho(\eta)_g$ in the large-$N$ limit (classical limit),
which is given by
\begin{equation}\label{rhog}
\rho(\eta)_g =  \left\{ \begin{array}{cc} 0 \, , & ~~~~~ 0\leq\eta\leq 1/2 \, ,  \\
\frac{(2\eta-1)}{2\eta} \, , &  \eta > 1/2 \,  .\end{array} \right.
\end{equation}
The results indicate that the condition $\langle\hat{n}_b^l/N\rangle\rightarrow0$ is strictly established in the
large $N$ limit for $\eta\in[0,1/2]$. Moreover, the critical feature in the $\rho(\eta)_g$ evolution shown in Fig.~\ref{F2} indicates that those presented in Fig.~\ref{F1} are indeed the finite-$N$ precursors of the U($n$)-SO($n+1$) QPT defined in the classical limit, which in turn confirms the consistency between the quantal analysis and classical analysis. It should be mentioned that one may get the similar results using other mean-field techniques such as the coherent state method adopted in \cite{Cejnar2007}, where a classical analysis of the phase structure of the interacting boson models in arbitrary dimension was given.

\begin{center}
\vskip.2cm\textbf{V. An example of the E(5) DS}
\end{center}\vskip.2cm

As is known, the concept of CPS was proposed~\cite{Iachello2000,Iachello2001} in the framework of Bohr-Mottelson model with the model predictions being well recognized in experiments~\cite{Casten2000,Casten2001,Clark2003,Clark2004}. Theoretically, the CPS method was extensively developed and became a "standard" way in modeling transitional structures of even-even nuclei~\cite{Iachello2003,Bonatsos2004,Bonatsos2005,Bonatsos2006,Bonatsos2004II,Caprio2002,Caprio2004,
Fortunato2004,Fortunato2006,Pietralla2004,Zhang2015,Zhang2017,Budaca2016,Budaca2018}, odd-A nuclei~\cite{Iachello2005,Alonso2007,Zhang2010I,Zhang2011I,Zhang2011II,Zhang2012I,Zhang2019I} and odd-odd nuclei~\cite{Zhang2021}. The relevant case in this work is the E(5) CPS~\cite{Iachello2000}. Compared with the differential realization~\cite{Iachello2000,Caprio2007}, the algebraic realization of the E(5) CPS was provided in \cite{ZLPSD2014,Zhang2014} and is currently generalized to more general cases. Note that the algebraic version of the E(5) CPS is just the E(5) DS. To give a concrete application of the E(5) DS, we adopt the Hamiltonian~\cite{Zhang2014}
\begin{equation}\label{HE(5)}\hat{H}_{\rm E(5)}=a~
\hat{C}_2[\mathrm{E}(5)]+b~\hat{C}_2[\mathrm{SO}(5)]+c~\hat{C}_2[\mathrm{SO}(3)]\,
\end{equation}
with $a$,~$b$, and $c$ being adjustable parameters. To solve the E(5) Hamiltonian, one can expand the eigenstates as~\cite{Zhang2014}
\begin{equation}\label{WF}
|\xi \,\tau \; \Delta L\rangle=\sum_{k=0}^m~C_k^\xi
(\hat{P}_{l=2}^\dag)^k |\tau \; \Delta L\rangle\, ,
\end{equation}
where $C_k^\xi$ are the expansion coefficients with $\xi$ denoting the
additional quantum number used to distinguish different states
with the same $\tau$, $\Delta$, and $L$ values. Here, $\tau,~\Delta$ and $L$ are the quantum numbers
as same as those in the U(5) basis defined in (\ref{u5}). In principle,
the dimension of the model space should be set to infinity due to the non-compactness of the E(5) group, which means that the number $m$
could be taken as $m\rightarrow\infty$. Nevertheless, a nice scaling behavior of $\mathrm{\hat{C}_2[E(5)]}$
has been revealed in ~\cite{ZLPSD2014}. The results suggest that a large enough value of $m$ can guarantee the numerical solutions rather accurately. Besides, it would be convenient to recale the parameter $a$ in (\ref{HE(5)}) with $a=\alpha m$ due to this scaling feature. More discussions on the solutions of the E(5) DS and their connections to the infinite square well potential adopted in the original E(5) CPS can be found in \cite{ZLPSD2014,Zhang2014}.

\begin{figure}
\begin{center}
\includegraphics[scale=0.35]{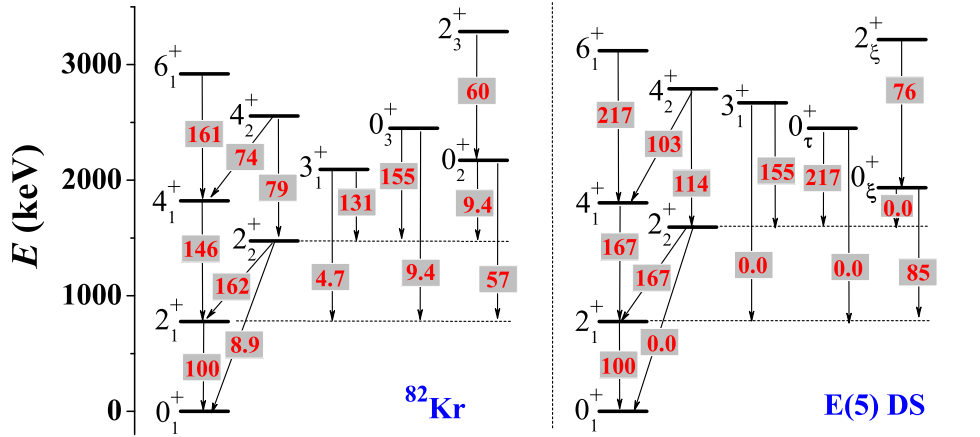}
\caption{The low-lying structure of $^{82}$Kr with the data taken from \cite{Rajbanshi2021} is shown to compare with
 the results solved from the E(5) Hamiltonian in (\ref{HE(5)}) with the truncation $m=200$, $\alpha=100.2$keV, $b=10$keV and $c=15$keV. In the comparison, all the $B(E2)$ results have been normalized to $B(E2;2_1^+\rightarrow0_1^+)=100$ (in any units), and $0_\tau^+$ and $0_\xi^+$ ($2_\xi^+$) represent the ones in theory with $\xi=1$ and $\xi=2$, respectively.
 \label{F3}}
\end{center}
\end{figure}

In experiments, $^{82}$Kr can be taken as an example of the E(5) DS, as this nucleus was very recently suggested to be a candidate of the E(5) CPS for the U(5)-SO(6) shape phase transition~\cite{Rajbanshi2021}. In Fig.~\ref{F3}, the low-lying level pattern of $^{82}$Kr is shown to compare with the results solved from the Hamiltonian (\ref{HE(5)}). For the $B(E2)$ transitions, the $E2$ transitional operator in theory is chosen as $\hat{T}_u^{\mathrm{E2}}=e(d^\dag+\tilde{d})_u^{(2)}$ with
the effective charge $e$ being determined by fitting the data for $B(E2;2_1^+\rightarrow0_1^+)$. Notably, the relative strength of $B(E2)$ transitions are only determined by the types of symmetry~\cite{IachelloBook87}, which means that the $B(E2)$ ratios in the E(5) DS should be independent of the parameters $a,~b,~c$ in the Hamiltonian. This point has been confirmed in the concretely numerical calculations~\cite{ZLPSD2014,Zhang2014}.
One can find from
Fig.~\ref{F3} that the experimental data for $^{82}$Kr are in good agreement with the results solved from the E(5) DS Hamiltonian (\ref{HE(5)}) for both the level energies and the $B(E2)$ structure.
Besides those in the $\xi=1$ family, the states in the $\xi=2$ family, $0_{\xi}^+$ and $2_{\xi}^+$, are also
well reproduced by the theoretical calculations. A small deviation from the experiments may be the $3_1^+$ state, to which the related
$B(E2)$ transitions are accurately predicted but the excitation energy is overestimated in theory by about 0.5 MeV.
It should be noted that the original E(5) CPS results including nearly the same $B(E2)$ structures as that shown in Fig.~\ref{F3}
can be also applied to compare with the experimental data~\cite{Rajbanshi2021}. However, no energy degeneracy appearing in experiments indicates that the SO(5) symmetry is evidently broken in this nucleus. Removing the energy degeneracies in the E(5) mode can be naturally realized by the present Hamiltonian (\ref{HE(5)}). Anyway, the results confirm that the the low-lying dynamics in $^{82}$Kr are indeed dominated by the E(5) DS~\cite{Rajbanshi2021}.

\begin{center}
\vskip.2cm\textbf{VI. Summary}
\end{center}\vskip.2cm

In summary, a boson algebraic realization of the $n$-dimensional Euclidean group symmetry with $n=2l+1$ has been proposed,
by which the E($n$) DS hidden in the $2^l$-pole deformed system described by the U($n+1$) vibron model is revealed.
Along the group contraction, it is shown that the E($n$) algebra may be equivalent to the SO($n+1$) algebra in the large $\sigma$ limit.
More importantly, it has been justified that the E($n$) DS can dynamically emerge at the critical point of the U($n$)-SO($n+1$) QPT under the condition that is strictly established in the classical limit and becomes a reasonable approximation for finite $N$. This point provides a solid theoretical basis for the CPS role of E($n$) in describing this
second-order QPT. This present study meanwhile generalizes our previous understanding of the E(5) DS for nuclear structure~\cite{ZLPSD2014,Zhang2014}.
As a new test, an E(5) DS Hamiltonian is applied to reproduce the low-lying structure of $^{82}$Kr. A good agreement between the experimental data and theoretical calculations confirms that the low-lying dynamics in this nucleus are indeed dominated by the E(5) DS, therefore adding another empirical evidence of the Euclidean dynamical symmetry in experiments. It is worth mentioning that the E(3) DS~\cite{Zhang2008} and E(2) DS~\cite{Clark2006} were also applied to explore nuclear properties. The former is just a specific case of E($n$) with $n=2l+1$, but the latter with $n=\mathrm{even}$ cannot be directly derived from the present discussions. A boson realization of the E(2) algebras has been given in \cite{Zhang2010} through considering a two-dimensional vector boson. The present analysis of $n=\mathrm{odd}$ might be extended along this line to $n=\mathrm{even}$, which will be discussed elsewhere.

\bigskip

\begin{acknowledgments}
We wish to thank Prof. Feng Pan for stimulating this work and for fruitful discussions on the related topics.
Supports from the Natural Science Foundation of China (11875158) is acknowledged.
\end{acknowledgments}

%\newpage

%\section*{References}

\end{document}